\documentclass[twocolumn,pre,tightenlines,nofootinbib,amsmath,amssymb,amsfonts,superscriptaddress]{revtex4}
\usepackage{subeqnarray}
\usepackage{hyperref}
\usepackage{lipsum}
\usepackage{accents}
\usepackage{array}

\usepackage{array}       
\usepackage{tabularx}
\usepackage{siunitx}
\usepackage{dcolumn} 

\usepackage{xcolor}

\usepackage{graphicx}
\usepackage{bm}
\usepackage{slashed}

\newcommand{\be}{\begin{equation}}
\newcommand{\ee}{\end{equation}}
\newcommand{\bea}{\begin{eqnarray}}
\newcommand{\eea}{\end{eqnarray}}
\newcommand{\beas}{\begin{subeqnarray}}
\newcommand{\eeas}{\end{subeqnarray}}

\newcommand{\dd}{{\rm d}}
\newcommand{\gr}[1]{{\bm #1}}
\newcommand{\com}[1]{}

\include{RevtexAuthorYear}
\begin{document}
\title{QED corrections to the big-bang nucleosynthesis reaction rates}

\author{Cyril Pitrou}
\email[]{pitrou@iap.fr}
\affiliation{CNRS, UMR 7095, Institut d'Astrophysique de Paris, 98bis Bd Arago, 75014 Paris, France.}

\author{Maxim Pospelov}
\email[]{mpospelov@perimeterinstitute.ca}
\affiliation{Perimeter Institute for Theoretical Physics, Waterloo, ON  N2J 2W9, Canada;\\
Department of Physics and Astronomy, University of Victoria, Victoria, BC V8P 5C2, Canada.}

\begin{abstract}
We compute radiative corrections to nuclear reaction rates that determine the outcome of 
the Big-Bang Nucleosynthesis (BBN). Any nuclear reaction producing a photon with an energy above $2m_e$
must be supplemented by the corresponding reaction where the final state photon is
replaced by an electron-positron pair. We find that pair prodution
brings a typical $0.2 \%$ enhancement to photon emission rates,
resulting in a similar size corrections to elemental abundances. The exception is 
$^4{\rm He}$ abundance, which is insensitive to the small changes in the 
nuclear reaction rates. We also investigate the effect of vacuum polarisation on
the Coulomb barrier, which brings a small extra correction when reaction 
rates are extrapolated from the measured energies to the BBN Gamow peak energies.
\end{abstract}

\maketitle

\section{Introduction}

Precise values of nuclear reaction rates are sometimes required in astrophysics and cosmology. 
While most of the applications belong to the stellar nucleosynthesis and solar neutrino physics, 
the one distinct cosmological application where precise nuclear physics is required is the 
Big-bang Nucleosynthesis (BBN). Developments of the last two decades, with an
independent input of the baryon-to-photon ratio from the Cosmic Microwave Background (CMB) anisotropies, 
paired with advances in observational extraction of the deuterium and helium abundance 
\cite{Aghanim:2018eyx,Coo18,Ave15}, 
have contributed to the rise of the standard cosmological model of the Universe. Further progress in refinement of BBN may only come from the combination of advances in CMB, more precise observations of primordial abundances, and 
\%-level determination of nuclear rates.

Quantitatively, the BBN has entered an era of precision, since
errors of $1.6\%$  (at one standard deviation) and smaller on
  $^4{\rm He}$ \cite{Ave15,Izo14} have been claimed in recent years.
At the same time, one should be cautious recognizing that it is very difficult to correctly estimate 
all systematic errors in helium abundance measurements, hence such a
  precision must not to be taken at face value. Furthermore, current
measurements of deuterium abundances \cite{Coo18} have reached an unprecedented $1.2\%$  precision. The total yield of $^4{\rm He}$ is essentially set by the neutron-to-proton ratio at the time of BBN that is mainly controlled by the weak interaction rates, while the feedback from other elements is small. 
Many subtle effects, including radiative corrections, have been taken into
account in Refs. \cite{Dicus1982,Lopez1997,LopezTurner1998,BrownSawyer,Serpico:2004gx},  and reviewed in Ref.~\cite{PRIMAT}, to reach a $0.1\%$ theoretical
precision on the weak rates.  At the moment, the main source of
theoretical uncertainty in the weak rates is the neutron lifetime
which is used as a proxy to estimate the strength of the weak interactions. The
incomplete decoupling of neutrinos prior to the reheating of photons
by electrons-positrons annihilations leads to a slight
modification of the energy density in neutrinos~\cite{Dolgov1997,Mangano2001,Mangano2005,Grohs2015}, 
which in turn affects the neutron abundance via the modification of the Hubble expansion rate
and flavour-specific distortions of the overall neutrino abundance.

Only traces amounts of other elements are left at the end
of BBN, and their final abundances depend on the nuclear rates of
a limited network of reactions, typically a dozen. The most relevant nuclei produced
during BBN are deuterium (${\rm D}$), $^3{\rm He}$, $^3{\rm H}$, $^7{\rm Li}$ and  $^7{\rm
  Be}$\footnote{$^7{\rm
  Be}$ decays subsequently to $^7{\rm  Li}$, and $^3{\rm  H}$ to
$^3{\rm He}$. It is customary to report the final abundances of $^7{\rm
  Li}$ and $^3{\rm  He}$ as their direct production augmented by these
decay products.}.  Reaching a percent level predictions for key nuclear reactions is highly 
desirable in the context of precise BBN predictions for these elements. 
In most cases, direct measurement of the reaction rate in experiment is performed, 
with some key reaction rates now being known down to $\approx 5\%$ accuracy. 
A theoretical {\em ab-initio} approach has also been successful. For a long time, 
theory determination of $n+p \leftrightarrow {\rm D}+\gamma$ rate has been more accurate than the 
corresponding experimental measurement (see, {\em e.g.} Ref.~\cite{Ando2005} for the 
nuclear effective theory approach to this rate), and the corresponding errors are at a sub-\% percent level. 
A significant theoretical progress has also being achieved in more
complicated rates such as D+$p$ and D+$n$ fusion processes
\cite{Mar05}, ${\rm D}+{\rm D}$ reactions~\cite{Arai2011}, $^3{\rm
    H}+{\rm D}
\leftrightarrow n+ {}^4{\rm He}$ and  $^3{\rm He}+{\rm D}
\leftrightarrow p+ {}^4{\rm He}$ reactions~\cite{Navratil2012},
$^3{\rm He}+{}^4{\rm He} \leftrightarrow {}^7{\rm Be} + \gamma$ and
$^3{\rm H}+{}^4{\rm He} \leftrightarrow {}^7{\rm Li} + \gamma$ reactions~\cite{Neff2011,Dohet-Eraly2015}.

Until now, the radiative corrections to  nuclear rates have been ignored in the compilation of the BBN reaction networks.
The goal of the current paper is to give a quantitative assessment of radiative corrections to the nuclear BBN rates, 
and determine the resulting shift in the yield for the main elements. There are two types of contributions that we need to 
consider: {\em i.} Replacement of an on-shell photon with a lepton pair, {\em ii.} Virtual excitation of the electron-positron vacuum,
modifying the Coulomb interaction of nuclei. There are several simplifying conditions that would allow us to perform this evaluation
{\em without} having to treat unknown nuclear matrix elements. Such conditions include a possibility of making a reliable 
multipole expansion, as the wavelengths of real/virtual photons are much larger than the characteristic 
nuclear size, $\lambda_\gamma \gg R_N$. The second simplifying condition is that the kinetic energies 
of incoming particles participating in the 
reactions are much smaller than the characteristic Gamow energies, $E\ll E_G$ (or equivalently, $Z_1Z_2e^2/(\hbar v) \gg 1$, where $v$ is relative velocity and $Z_{1(2)}e$ are charges of the reactants), and all reactions occur in the Coulomb-suppressed regime.

An important class of BBN reactions  involves a on-shell photon, 
with energy $E_\gamma > 2m_e$. Then, there exists a reaction where this photon is
 virtual, and decays into a lepton
pair~\cite{Kroll1955,Landsberg1985}.  In  other words, instead of a reaction 
${\rm N}_1+{\rm N}_2\to {\rm N}_3+\gamma$, one could always have ${\rm N}_1+{\rm N}_2\to {\rm N}_3+e^+e^-$ reaction. 
Importantly, for reactions in which the cross section is measured by detection of the
real photon, this extra channel is necessarily ignored and the inclusion of the final state pair 
will raise the total ${\rm N}_1+{\rm N}_2\to {\rm N}_3$ rate. 
We shall evaluate in section \ref{SecPair} this correction for the rates
\beas 
n+p &\to& {\rm D}+\gamma\,,\slabel{npD}\\
p+{\rm D} &\to& {}^3{\rm He} + \gamma \,,\slabel{pDHe}\\
^3{\rm H} + {}^4{\rm He} &\to& ^7{\rm Li} + \gamma\,,\\
^3{\rm He} + {}^4{\rm He} &\to& ^7{\rm Be} + \gamma\,,\\
^3{\rm H}+p &\to& {}^4{\rm He} + \gamma\,.
\eeas
Notice that for $^7{\rm Be}+p \to {\rm ^8B} +\gamma$ reaction, not
relevant for BBN (but important for {\em e.g.} solar neutrino physics), the energy release is not sufficient to produce 
a pair. In principle, there are also various cross channels for these reactions, ${\rm N}_3 + e^+e^- \to 
{\rm N}_1+{\rm N}_2 $, and ${\rm N}_3 + e^\pm \to 
{\rm N}_1+{\rm N}_2+ e^\pm $. In practice, these are going to be less important than the electron-positron 
production reactions, because at most relevant temperatures, $T<100$\,keV, the abundance of electrons and positrons in the 
thermal bath is quite small. 

Insertion of electron-positron loop inside a virtual photon is another way the radiative corrections 
tend to manifest themselves. For a nuclear reaction among charged particles at $E\ll E_G$,
the Coulomb repulsion is very important.  In the static approximation, 
this radiative effect modifies the Coulomb potential to what is known as the
Uehling-Serber potential~\cite{ALS1988,Trautmann1991,Bahcall1994}.
This potential - rather than a simple Coulomb $r^{-1}$ form - should be used in calculating the penetration factors. 
If the cross sections are measured at exactly the same range of energies relevant for the BBN reactions
({\em i.e.} Gamow peak energies), then the effect of Uehling-Serber potential is already accounted for in the 
measured cross section. If, however, one needs to extrapolate nuclear rates in energy (from the energy of the measurement to the 
Gamow peak) {\em and} precision is required, such an extrapolation must be done using the expression for the 
penetration factors in the Uehling-Serber potential.  In
the following, we shall consider how this effect modifies the
penetration factor of the key reaction \eqref{pDHe} (section
\ref{SecVP}).

Note that in principle there are additional types of
radiative corrections. For instance bremsstrahlung from one of the
charged reacting nuclei may in principle be taken into
account. However, for a nucleus with non-relativistic kinetic energy $E_K$ and mass $M$,
such an effect is typically suppressed not only by the fine structure constant $\alpha_{\rm FS}$ 
but also by powers of $E_K/M$. Similar suppressions 
would apply to any additional ``structural" photons, emitted from the nuclear transitions inside a nuclear reaction. 
With typical kinetic energies (${\rm MeV}$ of sub-${\rm MeV}$) this can
be safely disregarded. 

Finally, the implications for the freeze-out elemental abundances at the end of BBN are given in
section \ref{SecBBN}.

\section{Pair production reactions}\label{SecPair}

\subsection{Notation}

We review in detail the pair production corrections to reaction
\eqref{npD}, because we want to compare the relative importance of the
photon and the pair producing reactions
\beas
n+ p &\rightarrow& {\rm D}+ \gamma\slabel{GammaReaction}\,,\\
n+ p &\rightarrow& {\rm D}+ e^+ + e^-\slabel{EEreaction}\,.
\eeas
Throughout we use the particle physics metric signature
$(+,---)$. We denote the electron and positron momenta $p_1$ and $p_2$,
the deuterium momentum is $P'$, and the sum of the initial neutron and
proton momenta is $P$.  The electron, proton, neutron and deuterium
masses are respectively $m$, $m_p$, $m_n$, $m_{\rm D}$. The average nucleon
mass is taken to be  $m_N \equiv (m_n+m_p)/2$. $k$ is either the
photon momentum or the sum of the electron and positron momenta ($k =
p_1 + p_2$), depending on the reaction considered, such that in all cases from momentum conservation on has $P
= P' + k$. Finally,  $\epsilon^\mu$ is the photon polarization vector. Throughout our 
calculations we put $\hbar$ and $c$ to one.

\subsection{$E$-type and $B$-type  transitions}

We work in the Coulomb gauge for which the vector potential is only
spatial for a reference observer whose four-velocity is $u$, that is
$\epsilon \cdot u= 0$. In the rest frame of the reference observer,
$u^\mu = \delta_0^\mu$ and we split the photon momentum (be it real or virtual)
into frequency and spatial momentum as $k^\mu = (\omega, q^i)$, that
is $k \cdot u = \omega$. In practice, we use the center-of-mass rest
frame to define the reference observer. We further define the magnitude of the photon wave-vector $q$ and its
direction $n^i$, that is $q^i = q n^i$ with $\gr{n} \cdot
\gr{n}=1$. 
The transition matrix of the $\gamma$ reaction is necessarily of the form
\be
{\cal M} ={\cal J}_\mu \epsilon^\mu = {\cal J}_i \epsilon^i \,,
\ee
where ${\cal J}_\mu \equiv ({\cal J}_0, {\cal J}_i)$ is the nuclear
transition current. Given gauge invariance, it satisfies the transverse condition
\be\label{Magic0}
{\cal J}_\mu k^\mu =0\quad \Rightarrow \quad  {\cal J}_i q^i =  \omega {\cal J}_0\,.
\ee

We now consider that ${\cal J}_\mu$ can be expanded in powers of
$\omega/|\gr{p}_N|$ (or $q/|\gr{p}_N|$), where $\gr{p}_N$ is the typical
spatial momentum of the nucleons inside deuterium. This amounts to
assuming that the typical wavelength of photons $2\pi/\omega$ is much
larger than the typical size of the deuterium nucleus $\propto
1/|\gr{p}_N|$. Since $|\gr{p}_N| \approx \sqrt{2 m_{\rm D} B_{\rm D}}$,
where $B_{\rm D}$
is the deuterium binding energy, and the typical energy of the emitted
photon is $B_{\rm D}$, our expansion is in fact in powers of
$\sqrt{B_{\rm D}/(2m_{\rm D})} \approx 0.025$. At lowest order in this expansion,  ${\cal J}_i$ can
be separated into electric and magnetic dipole contributions as
\be
{\cal J}_i =  {\cal E}_i+  {\cal B}_i\,,
\ee
\beas\label{ModelEB}
{\cal E}_i &=& \omega d_i \,,\\
 {\cal B}_i &=& \epsilon_{jki} m^j q^k\,,
\eeas
in which $d_i$ (resp. $m_i$)  is the electric (resp. magnetic) dipole of
the nuclear transition.  Note that by construction ${\cal J}_0 = d_i q^i
$ and ${\cal B}_i q^i =0$. Eqs.~\eqref{ModelEB} can be recast in covariant form as
\beas\label{CovariantModel}
{\cal E}_\mu&=& - (k\cdot u) d_\mu + (d \cdot k) u_\mu \,,\\
{\cal B}_\beta &=& u^\mu m^\nu \epsilon_{\mu\nu\alpha \beta} k^\alpha\,.
\eeas
Physically, this is equivalent to considering either a coupling to the
Faraday tensor polarisation or its dual, since 
\beas
{\cal E}_\mu \epsilon^\mu &=& d_\mu u_\nu F^{\mu\nu}\,,\qquad F^{\mu\nu}  \equiv
k^{\mu} \epsilon^{\nu} -k^{\nu} \epsilon^{\mu}\,,\\
{\cal B}_\beta \epsilon^\beta  &=& \frac{1}{2}u^\mu m^\nu
\epsilon_{\mu\nu \alpha \beta} F^{\alpha\beta} = u^\mu m^\nu \widetilde{F}_{\mu\nu}\,.
\eeas

When averaging over all spins and directions, results can only depend
on the magnitude of momenta. Hence we can always replace
\be\label{AngularAverage}
d^\mu d^\nu \to -\frac{1}{3}|d|^2(g^{\mu\nu}-u^\mu u^\nu)
\ee
or $ d^i d^j \to |d|^2 \delta^{ij}/3$, where $|d|^2 \equiv - (d \cdot
d)$. We use a similar property and definitions for the magnetic dipole.
We use this arrow notation whenever the expression is reduced using this angular average.

\subsection{Photon producing reaction}

If we now compute $|{\cal M}|^2$ for the photon process, we get 
\be
|{\cal M}^2|_{(\gamma)}=\sum_{s=\pm1} {\cal J}_\mu {\cal J}^\star_{\nu} \epsilon_s^\mu \epsilon_s^{\nu\star} =
-g^{\mu\nu} {\cal J}_\mu {\cal J}_{\nu}^\star\,.
\ee
With minimal algebra we find
\beas
|{\cal M}|^2_{(\gamma),{\cal E}} &\to& \frac{2|d|^2}{3} |\gr{q}|^2\,,\\
|{\cal M}|^2_{(\gamma),{\cal B}} &\to& \frac{2|m|^2}{3} |\gr{q}|^2\,,
\eeas
noting that the parity considerations prevent interference terms in the angularly averaged  matrix elements.

\subsection{Pair producing reaction}

For the pair producing reaction, we treat ${\cal J}_\mu A^\mu$ as a vertex inside usual Feynman rules,
\be
|{\cal M}|^2_{(ee)} = {\cal J}_\mu{\cal J}^\star_\nu T^{\mu\nu}
\ee
where
\bea
T^{\mu\nu} &\equiv& \frac{e^2}{(k \cdot k )^2} (\slashed{p}_1 +m_e )\gamma^\mu (\slashed{p}_2 -m_e )\gamma^\nu\,,\nonumber\\
 &=& \frac{e^2}{k^4}\left(-g^{\mu\nu} 2 k^2 + 4 p_1^\mu p_2^\nu + 4 p_2^\mu p_1^\nu\right)\,.
\eea
Defining $E_1 = p_1 \cdot u$ and  $E_2 = p_2 \cdot u$ and using
angular averaging rules \eqref{AngularAverage}, we get 
\beas 
|{\cal M}|^2_{(ee),\cal E} 
&\to&\frac{2|d|^2 e^2}{3 k^4}\left[(E_1+E_2)^2(4 m_e^2+2 k^2)\right.\nonumber\\
&&\qquad \quad\left.-4 k^2 E_1 E_2+k^4\right]\,,\slabel{MeeE}\\
|{\cal M}|^2_{(ee),\cal B} &\to&\frac{2|m|^2 e^2}{3 k^4}\left\{4 m_e^2[(E_1+E_2)^2-k^2]\right.\nonumber\\
&&\left.\,\,-k^4 +2 k^2[(E_1)^2+(E_2)^2]\right\}\,. \slabel{MeeB}
\eeas

\subsection{Ratio of rates}

Ultimately, the cross sections of $n+p\to{\rm D}$ fusion are corrected by 
\be
\sigma^{(\gamma)} \to \sigma^{(\gamma)} (1+r)
\ee
where the ratio of the pair producing and photon producing rates is found from
\be\label{DefRatio}
r\equiv \frac{\sigma^{(ee)}}{\sigma^{(\gamma)}}  = \frac{\int \dd^5
  \Phi_{12\rm D}}{\int \dd^2 \Phi_{\gamma \rm D}} \,,
\ee
and is a function of reaction energy $E_K$. 
The phase-space elements are 
\bea
\dd^5 \Phi_{12 \rm D} &\equiv& (2\pi)^4\delta^4(P-k-P') |{\cal M}|^2_{(ee)} [\dd p_1]  [\dd p_2]  [\dd P'],\nonumber\\
\dd^2 \Phi_{\gamma \rm D} &\equiv& (2\pi)^4\delta^4(P-k-P')|{\cal M}|^2_{(\gamma)} [\dd p_\gamma]  [\dd P'],\nonumber
\eea
with $ [\dd p] \equiv \dd^3 \gr{p}/(2 E_\gr{p})$. Since the electric
and magnetic dipoles contribute separately, there is a ratio as defined in \eqref{DefRatio} for each case.

We choose to work in the center-of-mass frame in which $P^\mu =
E u^\mu$, where $E$ is the total initial energy. 
Furthermore, we also use an infinite deuterium mass expansion. This is
equivalent to expanding the results in a power series of $1/m_{\rm D}$, by
expressing all energies with respect to the deuterium mass, and then
considering only the lowest order term in such expansion. Physically,
it corresponds to ignoring deuterium recoil. Hence, in the photon producing case, the available energy $E_a$ goes entirely in the emitted
photon and we can use
\be\label{EnergyOmega}
\omega  = q = E_a \equiv E-m_{\rm D} = E_K + B_{\rm D}\,.
\ee
where $E_K \equiv E-m_n-m_p$ is the center-of-mass kinetic energy of
the initial neutron-proton pair.

The integral over phase space for the reference photon reaction is then simply
\be
\int \dd^2 \Phi_{\gamma \rm D}  = \frac{|{\cal M}|^2_{(\gamma)} \omega}{4\pi m_{\rm D}}\,.
\ee
As for the pair producing reactions we find
\bea\label{d2Phi12D}
\dd^2 \Phi_{12\rm D} &\equiv& \int _{\rm angles} \dd^5 \Phi_{12\rm D}\\
&=&\frac{|{\cal M}|^2_{(ee)}}{4 (2\pi)^3 (2E)^2} \dd (k^2) \dd (m^2_{2\rm D} )\,.\nonumber
\eea
where we used the invariant mass $m^2_{2 \rm D} \equiv (p_2 + P')^2$. Integrating over the range of allowed values for this invariant
mass, 
we find the differential ratios
\beas
\frac{\dd r_{\cal E}}{\dd k^2}&=& \left(1+\frac{x}{2}\right)\frac{\alpha_{\rm FS}}{6
  \pi} \sqrt{(1-x)(x-a)}(a+2x)\\
\frac{\dd r_{\cal B}}{\dd k^2}&=& (1-x)\frac{\alpha_{\rm FS}}{6\pi} \sqrt{(1-x)(x-a)}(a+2x)
\eeas
where
\be\label{PospelovNotation}
x \equiv k^2/E_a^2 \,,\qquad a\equiv 4 m_e^2/E_a^2\,.
\ee
Further integrating on the allowed values $4 m_e^2 \leq k^2 \leq E_a^2$ leads to 
\beas\label{RatioEBGood}
r_{\cal E} &=&\frac{\alpha_{\rm FS}}{9 \pi
  \sqrt{a}}\left[-a(10+a)E\left(1-a^{-1}\right)\right.\nonumber\\
&&\left.\qquad\qquad+(6 + \tfrac{7}{2} a+\tfrac{3}{2}a^2)K(1-a^{-1})\right],\slabel{RatioEGood}\\
r_{\cal B} &=&\frac{\alpha_{\rm FS}}{9 \pi
  \sqrt{a}}\left[a(-13+5 a)E\left(1-a^{-1}\right)\right.\nonumber\\
&&\left.\qquad\qquad+(6 + 5 a-3a^2)K(1-a^{-1})\right].\slabel{RatioBGood}
\eeas
In these expressions, $E$ is the complete elliptic integral of the
second kind, and $K$  is the complete elliptic integral of the first kind\footnote{These integrals are defined by $E(m) = \int_0^{\pi/2} (1-m \sin^2 \theta)^{1/2}\dd  \theta$ and $K(m) = \int_0^{\pi/2} (1-m \sin^2 \theta)^{-1/2}\dd
  \theta$.}.

\subsection{Analytic approximation of ratios}

In addition to the infinite deuterium mass approximation, one can also
consider a limit of energy release being much larger than the electron mass, $E_a\gg m_e$.  Expanding expressions from 
previous subsections in small $2m_e/E_a$, we arrive at 
\beas\label{RatioEB}
r_{\cal E} &\approx& \frac{2 \alpha_{\rm FS}}{3
  \pi}\left[\ln(2 E_a/m_e)-\frac{5}{3}\right]\,,\slabel{RatioE}\\
r_{\cal B} &\approx& \frac{2 \alpha_{\rm FS}}{3
  \pi}\left[\ln(2 E_a/m_e)-\frac{13}{6}+\frac{9 m_e^2}{2 E_a^2}\right]\,.\slabel{RatioB}
\eeas

\begin{figure}[htb!]
    \includegraphics[width=\columnwidth,angle=0]{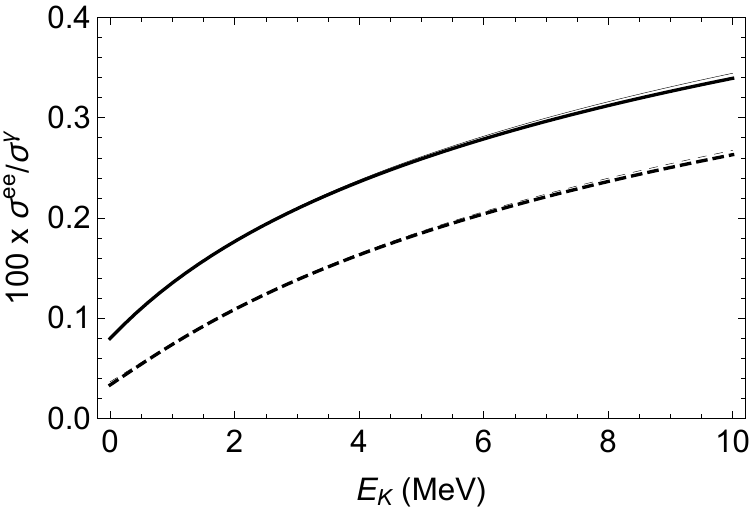}
     \caption{Solid is electric dipole ratio, dashed is magnetic dipole
       ratio. The result of the full numerical integration of
       \eqref{DefRatio} without approximation is depicted in thick lines, and the analytic approximations
       \eqref{RatioE} and \eqref{RatioB}  are in thin lines.}
\label{fig1} 
  \end{figure}

These approximations, along with the result of the direct numerical
integration of \eqref{DefRatio} without using the infinite deuterium
mass approximation, are depicted in Fig. \ref{fig1}. Note that the leading logarithmic terms in 
(\ref{RatioE}) and (\ref{RatioB}) has the same coefficient. It is model-independent, because the answer 
for the pair-production in this limit is dominated by a quasi-real photon.

\subsection{Comparison with literature}

Our results can be compared with the approach of \cite{Kroll1955}. However note that in this reference, a factor 2
is missing in front of the last term of Eq.~(6), hence there is a missing
factor $2$ in front of the $R_L$ term in its Eq.~(9). The validity of
Eq.~(9) in \cite{Kroll1955} is restored if one alters the definition of $R_L$ by adding an extra factor 2. 

With this convention, we follow \cite{Kroll1955} in defining the transverse and longitudinal components
of the ratios as
\beas
 R_T &\equiv& \frac{\int \dd^2 \Omega_{\hat{k}} {\cal J}^{(ee)}_i
  {\cal J}^{(ee)}_j (\delta^{ij}-\hat k^i \hat k^j)}{\int \dd^2 \Omega_{\gr{n}} {\cal J}^{(\gamma)}_i
 {\cal J}^{(\gamma)}_j (\delta^{ij}-n^i n^j)}\\
 R_L &\equiv& 2 \frac{\int \dd^2 \Omega_{\hat{k}} {\cal J}^{(ee)}_i
  {\cal J}^{(ee)}_j \hat k^i \hat k^j}{\int \dd^2 \Omega_{\gr{n}} {\cal J}^{(\gamma)}_i
  {\cal J}^{(\gamma)}_j (\delta^{ij}-n^i n^j)}
\eeas
where $\hat k^i$ is the unit vector in the direction of the spatial momentum of the virtual photon.

Our modeling \eqref{CovariantModel} of the electric and magnetic dipoles corresponds in the infinite deuterium mass approximation to
\begin{align}\label{rhochoices}
R_T^{\cal E} &= 1\,,&R_L^{\cal E} &= 1\,,\\
R_T^{\cal B} &=
        \left(\frac{|\gr{p}_1+\gr{p}_2|}{\omega}\right)^2\,,&R_L^{\cal
                                                              B}&= 0\,,
\end{align}
and from kinematics $|\gr{p}_1+\gr{p}_2|/\omega = \sqrt{1-x}$ in the notation \eqref{PospelovNotation}.

It is found in Eq.~(10) of \cite{Kroll1955}, that the subleading constant for $R_T=1$ and $R_L=0$ is $-11/6$ [in units of $2\alpha_{\rm
FS}/(3\pi)$]. Since our electric dipole parameterisation contributes to $R^{\cal E}_L=1$, it brings a subleading constant
contribution, hence the slightly different constant $-5/3$ found in \eqref{RatioE}. Furthermore the subleading constant in \eqref{RatioB}
is also different from $-11/6$, even though $R_L^{\cal B}=0$, due to the
non-constancy of $R_T^{\cal B}$ in our magnetic dipole parameterisation. 

Note that more generally, if  $R_T = C_T (|\gr{p}_1+\gr{p}_2|/\omega)^p$ and $R_L =
C_L (|\gr{p}_1+\gr{p}_2|/\omega)^q$, the differential ratio is
\bea
\frac{\dd r}{\dd k^2}&=&
\left[C_T (1-x)^{p/2}+C_L\frac{x}{2}(1-x)^{q/2}\right]\\
&&\times\frac{\alpha_{\rm FS}}{6\pi} \sqrt{(1-x)(x-a)}(a+2x)\,.\nonumber
\eea

\subsection{Reference rates}

It is well known that in the $n+p\to {\rm D}+\gamma$ reaction, at low energy the contribution of the magnetic 
channel is enhanced due to the virtual state close to threshold. 
Therefore, in order to know what the absolute correction of the BBN rates due to pair-production is, one 
needs to know how $\sigma_{(\gamma)}$ splits into the electric and magnetic components. While this calculation
has been treated in considerable detail over the years, we would use a simplified approach of ``zero range approximation" 
for the nuclear potential (see {\em e.g.} section 58 of \cite{LL4}), 
that results in $\sigma_{(\gamma)}$ within 10\% from a more accurate answer.

Photo-dissociation rates are the sum of electric and magnetic dipole contributions
\be
\sigma^{{\rm D}+\gamma \to n+p} = \sigma^{{\rm D}+\gamma \to n+p}_{\cal E}
+\sigma^{{\rm D}+\gamma \to n+p}_{\cal B}
\ee
where [Eqs. 58.4-58.7 of \cite{LL4}]
\beas\label{PhotoDissociation}
\sigma^{{\rm D}+\gamma \to n+p}_{\cal E} &=& \frac{8\pi \alpha_{\rm
    FS}}{3 m_N}\frac{\sqrt{B_{\rm D}}(\omega-B_{\rm D})^{3/2}}{\omega^3}
\slabel{PhotoD1}\\\slabel{PhotoD2}
\sigma^{{\rm D}+\gamma \to n+p}_{\cal B} &=&
\frac{8\pi}{3}(\mu_p-\mu_n)^2\\
&\times& \frac{\sqrt{B_{\rm D}(\omega-B_{\rm D})}(\sqrt{B_{\rm
      D}}+\sqrt{E_{s=0}})^2}{\omega(\omega-B_{\rm D} + E_{s=0})}\nonumber
\eeas
with $E_{s=0} \approx 0.067 \,{\rm MeV}$ the energy of the unstable
spin zero state of deuterium. The proton and neutron magnetic moments
are $\mu_p = 2.793 \mu_N$ and $\mu_n = -1.913 \mu_N$ where the nuclear Bohr
magneton can be written as $\mu_N \equiv \sqrt{\alpha_{\rm FS}}/(2m_p)$.

From detailed balance, they are related to the corresponding fusion rates by 
\be\label{Fusion}
\frac{\sigma^{n+p \to {\rm D}+\gamma}_{{\cal E}/{\cal
      B}}}{\sigma^{{\rm D}+\gamma
    \to n+p}_{{\cal E}/{\cal B}}} = \frac{3
  \omega^2}{2 m_N (\omega- B_{\rm D})}\,.
\ee
The cross sections for the reactions with 
pair production are estimated using \eqref{RatioEGood}
and  \eqref{RatioBGood}:
\be\label{MultiplyRatesRatios}
\sigma^{(ee)}_{{\cal E}/{\cal B}} \approx r_{{\cal E}/{\cal B}} \times
\sigma^{n+p \to {\rm D}+\gamma}_{{\cal E}/{\cal B} }\,.
\ee

\subsection{Average over thermal distributions}

In a thermalised plasma, the cross section must be averaged over the
thermal distribution of relative velocities between neutrons and protons in order to obtain interaction rates. The general expression is
\be\label{AverageT}
\langle \sigma v \rangle = \frac{2}{\sqrt{\pi T^3}}\int \sigma(E_K)
v_{\rm rel}(E_K) {\rm e}^{-E_K/T} \sqrt{E_K} \dd E_K
\ee
where $v_{\rm rel}(E_K) = \sqrt{2 E_K/\mu}$, with $\mu$ the reduced
mass of the reacting nuclei which for neutrons and protons forming
deuterium is approximately $m_N/2$.  

\begin{figure}[htb!]
    \includegraphics[width=\columnwidth,angle=0]{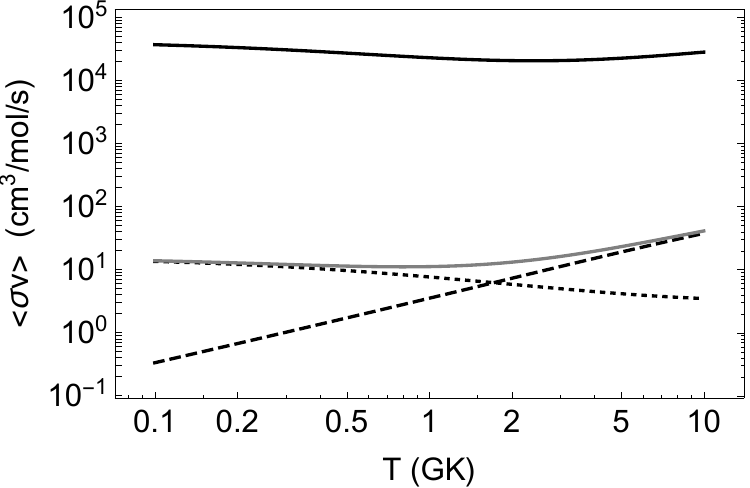}
     \caption{The electric (resp. magnetic) dipole contribution to $\langle \sigma v\rangle^{(ee)}$ is
       in dashed line (resp. dotted line). The  total correction is shown by the gray line while the reference value for
       $\langle \sigma v\rangle^{(\gamma)}$ (based on Eqs. (\ref{PhotoDissociation})) is in solid.}
\label{fig4} 
  \end{figure}

Performing the thermal average of reference rate $\langle\sigma v \rangle^{(\gamma)}$ by 
applying (\ref{AverageT}) to Eqs. (\ref{PhotoD1}) and  (\ref{PhotoD2}), we observe 
a rather good agreement with results of Ref.~\cite{Ando2005}, especially at low energies.
Combining the reference cross section [\eqref{Fusion} with \eqref{PhotoDissociation}] with the corrective
factors \eqref{RatioEBGood} as specified by
\eqref{MultiplyRatesRatios}, we are in position to estimate the magnitude of the
pair production corrections to the rates. The results are depicted in
Fig.~\ref{fig4}. A fitting formula in the range $0.05 <T_9 < 5$ is
\be
\frac{\langle\sigma v \rangle^{(\gamma)}+\langle\sigma v
  \rangle^{(ee)}}{\langle\sigma v \rangle^{(\gamma)}} = 1+\sum_{i=0}^4
a_i T_9^i
\ee
where $T_9$ is the temperature in ${\rm GK}$ and $a_0=3.6367\times10^{-4}$, $a_1=9.0830\times10^{-5}$, $a_2=4.1675\times10^{-5}$, $a_3=-1.1675\times10^{-5}$ and $a_4=1.0263\times10^{-6}$.

  It is easy to see that, at high temperatures, the relative size of the correction is sub-\%, while at lower temperatures, where 
  magnetic transition dominates, it drops to below the $10^{-3}$ level.

\subsection{Pair production for other reactions}\label{SecOther}

For all other reactions, the magnetic dipole contribution is
subdominant. However, for the $p+{\rm D} \to {}^3{\rm He} +
\gamma$ reaction, the magnetic dipole contribution is only marginally
subdominant since it contributes around $15\%$ at BBN energies (see
table 3 of Ref.~\cite{Schmid1997}). Given the size of the corrections
reported in table~\ref{TableCorrections} and the similarities between
the electric and dipole corrections [Eqs.~\eqref{RatioEB}] when the energy release exceeds $Q\gg 2m_e$, we ignore this
detail which would only bring a small correction to the already small correction. 
Hence, knowing the cross section for the photon emission
reaction, we just need to multiply it by the electric dipole corrective
factor~\eqref{RatioEGood} to get to the pair-production rate. 
An important subtlety is that the pair-production correction needs to be applied to rates
that had their cross sections measured via the detection of photon. 
The cross section of the reaction of $p+{\rm D} \to {}^3{\rm He} +
\gamma$ is a prime example of that, because it is measured via the final photon~\cite{Schmid1995,Ma1997,Cas02}. 
This also applies to reaction ${}^3{\rm H}+p
\to {}^4{\rm He} + \gamma$ in Ref.~\cite{Canon2002}, and to the most precise measurements of 
${}^3{\rm H} + {}^4{\rm He} \to {}^7{\rm Li} + \gamma$ reaction~\cite{Brune1994}. 

On the other hand, recommended reaction rates for the ${}^3{\rm He} + {}^4{\rm He} \to {}^7 {\rm Be} +
\gamma$ are based on the resulting  ${}^7 {\rm Be}$ activity in order
to keep statistics of the fits tractable~\cite{deBoer:2014hha,Adelberger2010}.
Therefore, the total rate is measured, including the radiative correction. Hence, we should not include the pair production correction to the reaction
  ${}^3{\rm He} + {}^4{\rm He} \to {}^7{\rm Be} + \gamma$.

Furthermore, since we are interested in corrections to the thermally averaged rates, one should in principle estimate the
correction to the thermally averaged rates by applying
\eqref{AverageT} to the corrected cross-sections. Given the size of the corrections, we find it sufficient to
multiply the thermally averaged rates of the photon producing
reactions by the corrective ratios~\eqref{RatioEGood} evaluated at the
most relevant kinetic energy for a given temperature and reaction ({\em i.e.} Gamow peak energy). For two nuclei
with $Z_1$ and $Z_2$ proton number and reduced mass $\mu = u A_1
A_2/(A_1+A_2)$, it is given (see Ref.~\cite{NACRE}) by 
\bea\label{MostLikely}
E_K^{\rm ML}(T_9) &\approx&
\left(\frac{\mu}{2}\right)^{1/3}\left(\pi  \alpha_{\rm FS} Z_1
    Z_2 k_B T\right)^{2/3} \nonumber\\
&\approx& 0.1220 (\mu/u)^{1/3}  (Z_1 Z_2 T_9)^{2/3}\,{\rm MeV}\,.
\eea

For the photon emission reactions relevant for BBN, the
corrections at $T_9 = 0.8$ are reported in
Table~\ref{TableCorrections}. Evidently it affects mostly reactions
with larger $Q$ values, since a larger available energy brings larger
values for the logarithms of Eq.~\eqref{RatioE}.

{\renewcommand{\arraystretch}{1.4}%
\begin{table}[htb!]
\begin{tabularx}{\columnwidth}{X|S[table-format=4.6]|S[table-format=7.8]}
  \hline
  Reaction & \multicolumn{1}{c|}{$Q$ in MeV} & \multicolumn{1}{c}{Correction (in $\%$)} \\
  \hline
  $p+{\rm D} \to {}^3{\rm He} + \gamma$ & 5.493 & 0.220 \\
  ${}^3{\rm H} + {}^4{\rm He} \to {}^7{\rm Li} + \gamma$ & 2.468 & 0.106 \\
  ${}^3{\rm He} + {}^4{\rm He} \to {}^7{\rm Be} + \gamma$ & 1.587 & 0.057 \\
  ${}^3{\rm H}+p \to {}^4{\rm He} + \gamma$ & 19.81 & 0.416\\
  \hline
\end{tabularx}
\caption{Correction from pair production to the thermally averaged rates, at $T_9=0.8$. }
\label{TableCorrections}
\end{table}}
Even the largest correction in this table is still below the current errors for the corresponding photon emission reaction rates.

\section{Vacuum polarization corrections}\label{SecVP}


Vacuum polarisation modifies the Coulomb potential, and this affects
how cross-sections are extrapolated in energy 
when taking into account the Gamow
penetration factor. We find that this is relevant for the reaction
\eqref{pDHe} and we detail in this section how and when this must be
taken into account. We follow Ref.~\cite{Bahcall1994} which applied these
corrections in the context of solar nuclear reactions. 

Let us consider two charged nuclei $1$ and $2$ with charges $Z_1$ and $Z_2$ and reduced mass
$\mu$. The Coulomb potential is 
\be\label{VC}
V_C(r) = \frac{Z_1 Z_2 e^2}{4\pi r } = \frac{Z_1 Z_2 \alpha_{\rm FS}}{r }\,.
\ee
The Uehling-Serber potential~\cite{Serber:1935ui,Uehling1935} is an additive correction which takes
into account electron vacuum polarisation, that is fermionic loops in the
photon propagator. The total potential takes thus the form  $V= V_C + V_U$ with
\be
V_U(r) =V_C(r) \times \frac{2 \alpha_{\rm FS}}{3 \pi}I(r)\,.
\ee
The Uehling function is given by
\be
I(r) \equiv \int_1^\infty {\rm e}^{-2 m_e r
  x}\left(1+\frac{1}{2x^2}\right)\frac{\sqrt{x^2-1}}{x^2} \dd x\,.
\ee
Usually, nuclear reaction cross sections are proportional to the Gamow penetration factor related to the electrostatic potential, which is  
\be\label{DefGamma}
\Gamma(E) = \exp\left[-2\int_{r_{\rm min}}^b \sqrt{2\mu (V(r)-E)}
\right] \dd r\,,
\ee
where $b$ is the turning point defined by the classical bareer $V(b)
\equiv E$, that is $b = Z_1 Z_2 \alpha_{\rm FS}/E$, and $r_{\rm min}$
is the distance at which nuclear forces overcome the repulsive
barrier, typically the sum of the radii of the interacting nuclei.

When considering only the Coulomb potential~\eqref{VC} and setting $r_{\rm min}=0$, the penetration factor can be expressed with the Sommerfeld parameter
\be
\eta \equiv \frac{Z_1 Z_2 \alpha_{\rm FS}}{\sqrt{2E/\mu}}
\ee
since after performing the integral in \eqref{DefGamma} we get
\be
\Gamma(E) = \exp[-2\pi \eta]\,.
\ee
In Ref.~\cite{Bahcall1994} it is shown that taking into account the Uehling-Serber
potential translates into a  modification of the Gamow penetration
factor by
\be
\Gamma_{\rm C+U}(E) = \Gamma_{\rm C}(E)[1- \Delta(E)]\,,
\ee
where
\be\label{DefDelta}
\Delta(E) \equiv \frac{4 \alpha_{\rm FS} \eta}{3 \pi} \int_0^1 \frac{I(by)}{\sqrt{y-y^2}} \dd y\,.
\ee
The function $\Delta(E)$ is evaluated by switching the order of  integration.  The $y$-integral 
can be performed analytically, leaving us with one numerical integral over $x$,
\bea\label{FirstDelta}
\Delta(E) &=&\frac{4 \alpha_{\rm FS} \eta}{3 \pi}  \int_1^\infty
J\left(\frac{m_e x Z_1 Z_2 \alpha_{\rm FS}}{E}\right)\nonumber\\
&&\times\left(1+\frac{1}{2x^2}\right)\frac{\sqrt{x^2-1}}{x^2} \dd x\,,
\eea
with the definition in terms of a Bessel function of the first kind
$J(z) \equiv \pi I_0(z) \exp(-z)$. 

\begin{figure}[htb!]
    \includegraphics[width=\columnwidth,angle=0]{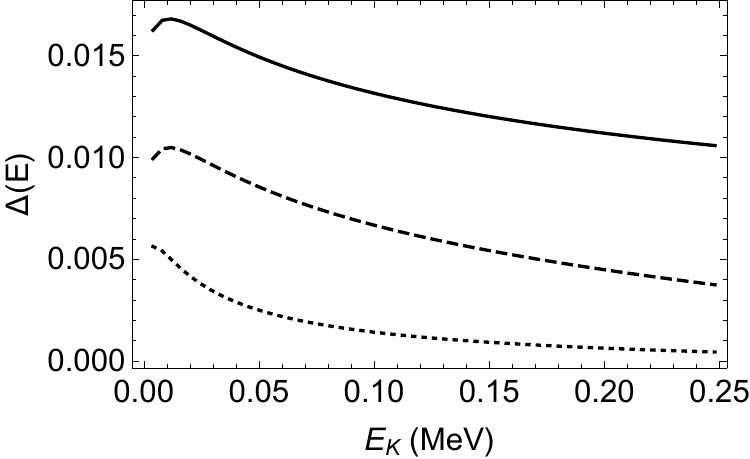}
     \caption{Relative variation of the Gamow penetration factor
       $\Delta(E)$ for the reaction \eqref{pDHe} when ignoring the
       centrifugal barrier ($r_{\min}=0$ in continuous line and
       $r_{\rm min}=3\, {\rm fm}$ in dashed line), and with $\ell=1$
       centrifugal (and also $r_{\rm min}=3\, {\rm fm}$) barrier in dotted line.}
\label{figDelta} 
\end{figure}

However, one cannot ignore the centrifugal barrier which adds to the
effective repulsive potential~\cite{Clayton1983}, if the interaction
takes place via non-zero orbital angular momentum waves. From Fig. 1
and table 3 of Ref.~\cite{Schmid1997}, one infers that it is precisely the case
that the $\ell=1$ wave accounts for approximately $85\%$ of the
reaction~\eqref{pDHe} at BBN energies. The centrifugal effective
potential is
\be\label{Vcen}
V_{\rm cen} = \frac{\ell(\ell+1)}{2 \mu r^2}\,,
\ee
where $\mu$ is the reduced mass of the two nuclei interacting. For
large nuclei, this extra contribution remains small compared with the pure
Coulomb barrier~\eqref{VC} and it is possible to treat it as a
perturbation as detailed in section 4.5 of Ref.~\cite{Clayton1983}. However this is not the case for the
reaction~\eqref{pDHe} at BBN energies and one must in principle
estimate the centrifugal barrier directly from $V_{C}+V_{\rm cen}$
replaced in \eqref{DefGamma}. The Uehling-Serber potential can still
be considered as a perturbation with respect to $V_{C}+V_{\rm cen}$,
so the correction to the penetration factor, when including the
centrifugal barrier, is given by
\be\label{DefDelta2}
\Delta(E) \equiv \frac{4 \alpha_{\rm FS} \eta}{3 \pi} \int_{\frac{r_{\rm
    min}}{b}}^1 \frac{I(by)}{\sqrt{y-y^2+\frac{\ell(\ell+1) E}{2\mu (Z_1 Z_2
    \alpha_{\rm FS})^2}}} \dd y\,.
\ee
It is not simple to reduce analytically this expression to a one
dimensional integral, as was the case without the centrifugal barrier,
and we must evaluate numerically directly this two-dimensional
integral. The corrections~\eqref{FirstDelta} and~\eqref{DefDelta2} for the reaction \eqref{pDHe} are
plotted on Fig.~\ref{figDelta}. In the following, we consider only the
$\ell=1$ penetration factor correction from \eqref{DefDelta2} (with
$r_{\rm min} \approx 3\,{\rm fm}$ which is roughly the sum of the nuclear radii), that is we ignore the order $15\%$ contribution of the magnetic dipole
transition in reaction~\eqref{pDHe} at BBN energies. 

Cross-sections that are evaluated at energies different from the most likely energy (the Gamow peak), must be extrapolated to the 
relevant energy range,
\be
\sigma_{\rm C+U}(E) = \sigma_{\rm C}(E) \frac{[1-\Delta(E)]}{[1-\Delta(E_{\rm meas})]}\,.
\ee
The most precise measurement of reaction  \eqref{pDHe} was performed at energies smaller than the BBN Gamow peak. 
Theoretical extrapolations to the BBN energy range was done with the method described in \cite{Coc15},  
which consists of using mainly the reliable low-energy data \cite{Cas02} and the theoretical model \cite{Mar05}.
  PRIMAT \cite{PRIMAT} uses the results of the most recent~\cite{Bayes16}
  which implements this type of approach with Bayesian statistics. The
  energy range of the main source of statistics comes from the
  measurements of \cite{Cas02} which span $3\,{\rm keV}<E_K<20\,{\rm
    keV}$. On Fig. \ref{figDelta} we note that this is precisely the
  range of values for which the function $\Delta(E)$ is nearly
  constant. Hence we choose to evaluate this function at the value  $E_{\rm meas} = 15\,{\rm keV}$.

\begin{figure}[htb!]
    \includegraphics[width=\columnwidth,angle=0]{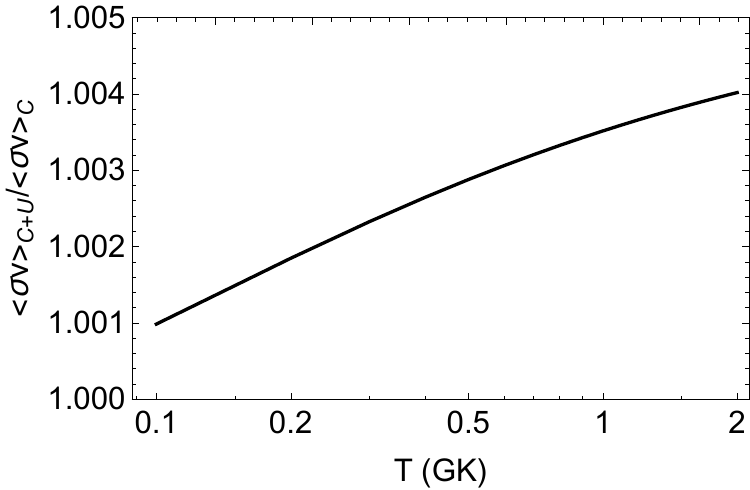}
     \caption{Corrective ratio $\langle [\sigma v]\rangle_{\rm C+U}
       /\langle [\sigma v]\rangle_{\rm C}$ for the reaction \eqref{pDHe}.}
\label{figDelta2} 
\end{figure}

To subsequently estimate how this correction affects the thermally
averaged rates of reaction \eqref{pDHe}, we can evaluate it at the most likely energy \eqref{MostLikely}, so that
\be\label{RatioSigmas}
\frac{\langle[\sigma v]\rangle_{\rm C+U} }{\langle[\sigma v]\rangle_{\rm C} } \approx 
\frac{[1-\Delta(E_{K}^{\rm ML}(T))]}{[1-\Delta(E_{\rm meas})]}
\approx 1 + \sum_{i=0}^4 b_i T_9^i\,,
\ee
where the last equality is a fit with $b_0 = -3.5264\times 10^{-4}$,
$b_1 = 1.4925\times 10^{-2}$, $b_2 = -2.6236\times 10^{-2}$, $b_3 =
2.3837\times 10^{-2}$, $b_4 = -1.0443\times 10^{-2}$, $b_5 = 1.7473\times 10^{-3}$ which is valid
for $0.05 \leq T_9 \leq 2$. The ratio \eqref{RatioSigmas} of the corrected to the uncorrected rates for reaction \eqref{pDHe} is plotted in Fig. \ref{figDelta2}.
Let us emphasize again that we have only performed a crude estimation
of the correction, since once we can use reliable data in the range $E_K = 100-200\,{\rm
  keV}$, as should be available in the future \cite{Gus17}, then this correction would no longer need to be taken into account.
Alternatively, if the theoretical extrapolation are still used, 
future precision calculations of the nuclear reaction rates could
include the Uehling-Serber correction into account directly in the asymptotics of the Coulomb wave functions. 

\section{Effect on BBN}\label{SecBBN}

We now review the effect of the corrections discussed in the previous
sections on elemental abundances at the end of BBN. For any photon 
emission reaction, the reverse rate is obtained using
detailed balance, as explained in  detail in section 4.2 of Ref.~\cite{PRIMAT}. This is
also the case for the pair-production reactions since electrons and
positrons are always at thermal equilibrium with photons during
BBN. Hence one can compute the sum of the photon and pair producing
reactions, and then obtain the sum of the backward rates from the very
same detailed balance relation. To summarise, if the forward rate is
increased, the backward rate is increased by the same amount so as to
always satisfy the detailed balance conditions.

Note that when considering reactions with electron-positron pairs, one
should also consider in principle $2\to 3$ reactions
\be
{\rm D}+e^\pm \leftrightarrow p+n + e^\pm\,.
\ee
These reactions turn out to be negligible for the following reason. 
The initial electron or positron needs to bring not only the necessary binding
energy, but also the final rest mass of the electron or positron. If a
photon needs only $E_\gamma = B_{\rm D}$ to dissociate deuterium, an
electron would need $B_{\rm D}+m_e$. Hence at energies $T \ll m_e$, from the thermal
equilibrium distribution of electrons, one deduces that the reaction is
suppressed by a factor $\exp(-m_e/T)$ compared with the photon
dissociation reaction. Since nucleosynthesis really starts only below
$T \approx 100 \,{\rm keV}$, this is an additional prohibitive suppression factor. The
argument can also be made for the reverse rate, and in that case the
suppression factor comes from the low probability of having the three
body reaction, since this brings an extra factor $\propto n_e /T^3 \propto
\exp(-m_e/T)$.

The effect on BBN of all corrections discussed in the previous is
summarised in Table~\ref{TableCorrections2}. Deuterium is by far the
most precisely measured component, and we note that in total its
predicted abundance is decreased by $0.19\,\%$.

{\renewcommand{\arraystretch}{1.4}%
\begin{table}[htb!]
\begin{tabular}{l|S[table-format=2.5]|S[table-format=2.5]|S[table-format=2.5]|c}
  \hline
  Element & \multicolumn{1}{c|}{Reference} & \multicolumn{1}{c|}{Pair}
  & \multicolumn{1}{c|}{Pair+Uehling} &  Tot. Var. $\%$ \\
 \hline
$10^5 \, {\rm D}/{\rm H}$ & 2.4585 &2.4564& 2.4539&$-0.19$\\
$10^5 \, {}^3{\rm He}/{\rm H}$ & 1.0742& 1.0752& 1.0763&$+0.20$\\
$10^{10} \, {}^7{\rm Li}/{\rm H}$ & 5.670& 5.684& 5.695& $+0.44$\\
  \hline
\end{tabular}
\caption{Variations of BBN abundances. The first column gives reference abundances
  computed with ${\rm PRIMAT}$ \cite{PRIMAT}. The second column takes
  into account the effect of pair production in photon producing processes, and
  the third column also includes the Uehling potential corrections to
  reaction \eqref{pDHe}.  Variation of ${}^4{\rm He}$ is insignificant.}
\label{TableCorrections2}
\end{table}
}

\section*{Conclusion}

Radiative corrections from pair production have a relative
size set by the prefactor $\alpha_{\rm FS}/\pi$. 
In the limit of the energy release being much larger than $2m_e$,
 the pair-production radiative correction is completely fixed by the 
photon emission cross section. In reality, this limit does not 
hold for most of the reactions, and we model the corresponding 
nuclear EM transition vertex by the simplest electric dipole form.
Most of the photon producing reaction rates are indeed dominated by electric dipole transitions,
and are easily corrected by a universal function. Synthesis of 
deuterium is a notable exception, where at lowest energies the reaction is dominated by a
magnetic transition. The reaction  $p+{\rm D} \to {}^3{\rm He} +
\gamma$ is also partially determined by the magnetic dipole transition at
BBN energies, but we ignored this complication as detailed in
section~\ref{SecOther}, because the relative difference between corrections to 
magnetic and electric transitions becomes less important as $Q$ is taken to be much larger than $2m_e$.  
In the limit of zero recoil, we give complete analytic expressions 
for the pair-production corrections induced by $E1$ and $M1$ nuclear transitions, and 
correct some mistakes (or typos) in the original reference \cite{Kroll1955}.
Finally, the Uehling-Serber potential 
creates a finite correction to the Gamow penetration factor, and it needs to be taken into 
account if the nuclear rates involve extrapolation in energy. 

Having evaluated the radiative corrections to nuclear rates, we determine the resulting change in the 
BBN predictions. The radiative-correction-induced shifts to freeze-out abundances 
are small, below the percent level, and are summarized in 
Table II. Nevertheless, expected progress 
in CMB physics, observations of deuterium, nuclear physics measurements of the relevant 
rates, as well as in {\em ab-initio} calculations of nuclear reactions, may make even sub-percent shifts relevant
in the future. The abundance of  $^4{\rm He}$ is very insensitive to the change in nuclear rates,
and thus the impact of radiative corrections on helium abundance is negligible.

We thank Alain Coc, Jean-Philippe Uzan and Elisabeth Vangioni for
discussions on the topic. CP thanks the Perimeter Institute (Waterloo, Ontario, Canada) and the
University of Victoria (British Columbia, Canada), where this work was
initiated for the former and completed for the latter. MP is grateful to the groups 
at IAP and Orsay for kind hospitality. Research at Perimeter Institute is supported by the
Government of Canada through Industry Canada and
by the Province of Ontario through the Ministry of Economic Development \& Innovation.



\bibliography{NuclearBib}

\end{document}